\documentclass[journal,twocolumn]{IEEEtran}

\usepackage[utf8]{inputenc}
\usepackage{color,graphics,graphicx}
\usepackage[cmex10]{amsmath}
\usepackage{amsfonts,amssymb,bm}
\usepackage{subfig}

\def\H{{\mathbf H}}
\def\I{{\mathbf I}}

\def\0{{\mathbf 0}}
\def\1{{\mathbf 1}}

\def\u{{\mathbf u}}

\def\w{{\mathbf w}}
\def\x{{\mathbf x}}
\def\y{{\mathbf y}}

\DeclareMathOperator{\trace}{tr}

\newcommand{\openone}{\leavevmode\hbox{\small1\normalsize\kern-.33em1}}

\newcommand{\snr}{\mathsf{snr}}
\newcommand{\inr}{\mathsf{inr}}

\newcommand{\ex}[1]{\mathbb{E}\left[ {#1}\right]}

\begin{document}

\title{Non-Coherent Rate Splitting for the MISO BC with Magnitude CSIT}

\author{ Carlos Mosquera,  Tom\'as Ram\'{\i}rez, M\`arius Caus, Nele Noels, Adriano Pastore\thanks{This work has been partially supported by European Space Agency funded activity SatNEx IV CoO2-Part 1 WI 3 "Non-Orthogonal Superposition Techniques for Multi-Beam Satellite Networks". The views of the authors of this paper do not reflect the views of ESA. Also funded by the  Agencia Estatal de Investigacion (Spain) and the European Regional Development Fund (ERDF) through the project  MYRADA (TEC2016-75103-C2-2-R).
    Carlos Mosquera and  Tom\'as Ram\'{\i}rez are with atlanTTic  research  center,  University of  Vigo, Galicia, Spain. (e-mail: \{mosquera, tramirez\}@gts.uvigo.es). M\`arius Caus  and  Adriano Pastore are with Centre Tecnol\`ogic de Telecomunicacions de Catalunya (CTTC), Castelldefels, Barcelona, Spain (e-mail:   \{marius.caus, adriano.pastore\}@cttc.cat). Nele Noels is with DIGCOM research group, TELIN Department, Ghent University, Ghent, Belgium  (e-mail: Nele.Noels@ugent.be).}}



\maketitle

\begin{abstract}
  A rate splitting based scheme is proposed to operate a broadcast setting with two antennas at the transmit side and two single-antenna receiving
  terminals.
  The transmitter knows the magnitude of the channel coefficients, and it is oblivious to the phase  information.  Each transmit antenna, with a power constraint, sends a private message and a common message to be decoded by both receivers.   An achievable rate region is obtained, which enlarges the capacity region of the vector broadcast channel with vector channel magnitude feedback by means of  superposition coding. 
\end{abstract}


\section{Introduction}

Under full Channel State Information at the Transmitter (CSIT), the Shannon capacity  region of the MIMO Broadcast Channel (BC)
is achieved by means of dirty-paper coding (DPC) \cite{ElGamal2011}. However, the information available to the transmitter could  be limited by different practical constraints,
mainly due to the overhead imposed on the communication itself by the training symbols and exchange of channel estimates between receiver and transmitter.
As a result, the transmitter has a partial knowledge of the channel in the best case, with finite precision and/or outdated estimates.
With partial CSIT, the computation of the Shannon capacity region has proven elusive, although major steps have been done recently by using, for example,
the degrees-of-freedom (DoF) \cite{ClerckxTCOM2016}  and generalized degrees-of-freedom (GDoF) \cite{JafarISIT2016} frameworks.
These are asymptotic metrics in the signal to noise ratio  which provide useful insights  for practical encoding mechanisms. 
For partial CSIT, when the  channel estimation error scales with the signal to noise ratio, both references have proposed schemes relying on rate splitting strategies, already considered by Han and Kobayashi in their seminal paper \cite{HanKobayashi1981} for the interference channel.

In the current work we will consider the two-user  two-antenna Multiple-Input Signal-Output   Broadcast Channel (MISO-BC).  
This setting, also known as the vector broadcast channel with single-antenna terminals \cite{jafar2005}, fits several practical scenarios such as cellular systems with multi-antenna base stations \cite{heath2018}
or the multibeam satellite downlink channel \cite{chatzinotas2015}, \cite{ramirez2018}.

Channel phase  usually varies faster than magnitude, and may pose a significant burden if reported  back to the transmitter \cite{payaro2005}. Thus,
departing from the finite precision CSIT problem investigated in \cite{ClerckxTCOM2016,JafarISIT2016},
we  will assume a channel with fixed magnitude for the duration of the codewords, and  phase unbeknown to the transmitter, which is only able 
to gain access to  the magnitude of the channel coefficients; on the other side,  the receivers are assumed to have perfect channel knowledge. 


We propose a particular rate splitting scheme to address this setting, with  a public  message  multicast to the receivers on top of the private messages from each antenna. As a Non-Orthogonal Multiple Access (NOMA) scheme,  is such that the achievable rates are independent of the channel phases, and the knowledge of channel magnitudes is exploited to optimize the relative weights of each message.
Numerical results for the two-user case will reveal how rate splitting enlarges the rate region of the baseline design based on superposition coding. With respect to \cite{ramirez2018},
where rate splitting was pointed out as a method to deal with the lack of phase information in a multibeam satellite setting, a more detailed study is performed here, including a suitable theoretical framework, a general overview of the relevant schemes, and the evaluation under different operation regimes.

After presenting the system model in Section \ref{sec:system}, a brief review of previous results is exposed in Section \ref{sec:review}. The main contribution of the paper is contained  in Section \ref{sec:NCRS}, with numerical results detailed in Section \ref{sec:results} before the conclusions.


\section{System Model}\label{sec:system}

For a generic number of users $N$, the  MISO-BC is  modeled as $\y = \H \x + \w$, with $\y \in \mathbb C^{N\times 1}$
the received values at the $N$ single-antenna user terminals, $\H \in \mathbb C^{N \times N}$ the square channel matrix, $\x  \in \mathbb C^{N\times 1}$
the symbol vector transmitted by  the $N$ antennas,  
$\w \in \mathbb C^{N\times 1}$ zero-mean unit variance Additive White Gaussian Noise (AWGN), such that $\ex{\w \w^H} = \I_N$.
The transmit power is given by $P = \trace\{\x \x^H\}$, and for commonly found technological constraints, the average per-antenna power will be upper bounded by  $P/N$.

  We consider a block-fading channel, with the magnitude of the channel entries $[\H]_{ij} = h_{ij}$ constant during the transmission of a codeword.
   We will  focus on the two-user MISO-BC,  whose system equation reads as 
\begin{gather}
  y_1 = h_{11} x_1 + h_{12} x_2 + w_1,\\
  y_2 = h_{21} x_1 + h_{22} x_2 + w_2.
\end{gather}
The transmitter knows the channel quality of all links:
\begin{equation}
\gamma_{ij} = \frac{P}{2} |h_{ij}|^2, 
\end{equation}
which represent signal to noise and interference to noise ratios, respectively, and thus shall alternatively be denoted as 
\begin{equation}\label{eq:snrs}
\snr_1 = \gamma_{11},\; \snr_2 = \gamma_{22}, \;  \inr_1 = \gamma_{12}, \; \inr_2 = \gamma_{21}.
\end{equation}
On the other side, the receivers have full knowledge of the channel entries $h_{ij}$.

We are interested in computing an achievable rate region of this channel,
when the transmitter  is oblivious to the channel phase information.
The corresponding setting without cooperation among the transmitters feeding the antennas is known as  the interference channel (IC).
The capacity region of the IC is unknown, although  the Han-Kobayashi (HK) achievable rate 
\cite{HanKobayashi1981} is a well-known inner bound; this scheme yields the best single-letter inner bound for the performance of the IC, and it is based on the splitting of the messages into a private and a common part (to be decoded by both receivers).
Next we recapitulate some of the existing results for the cooperative case, i.e., the MISO-BC.

\section{Review of Related Results}\label{sec:review}

The problem under consideration can be stated as the computation of an achievable rate region for the MISO-BC with perfect Channel State Information at the Receiver (CSIR), partial CSIT based on the feedback of the different channel magnitudes, and per-antenna power constraints.
This is a highly relevant setting in practice, as detailed above, with unknown capacity region. 
Next,  we review several  known results for scenarios  which share some (not all) of the features of the problem under study.

The capacity region of the MIMO-BC assuming full CSIT and global  transmit power constraint is achieved by Dirty Paper Coding (DPC) \cite{ElGamal2011},
a non-linear scheme with high implementation complexity.
Under per-antenna power constraints, the corresponding capacity region is necessarily smaller, and has been derived in \cite{yu2007}. Both cases, with  full CSIT required, are not applicable if phase information is not available. 

The vector broadcast channel with perfect CSIR and vector channel  magnitude information at the transmitter has been studied in \cite{jafar2005}. In this case the receivers report the total received power, which amounts to the transmitter having access to the norm of the different rows of the channel matrix $\H$. It is proved in \cite{jafar2005} that the capacity region is the same as that for the scalar broadcast case, and it is achieved by superposition coding.
From \cite[Lemma 5]{jafar2005}, the transmit vector $\x = \u_1+\ldots+\u_N$ is the superposition of $N$ independent jointly Gaussian vectors, each following  $\u_n \sim {\cal CN}(0, \I_N P_n/N)$, and $P_n$ the power assigned to the $n$th user. The boundary of this capacity region  will set an inner bound for our scheme, which can exploit the additional information provided by the individual weights of the different links.
The knowledge of the weight of individual paths provides some valuable information for the rate splitting, improving on the superposition coding as analyzed in the following section.  

Also related to the MISO BC under study is the MISO channel with one receive terminal and per-antenna power constraints; this case is analyzed in
\cite{vu2011}, including also the lack of CSIT.

\section{Non-Coherent Rate Splitting}\label{sec:NCRS}

\begin{figure}[!t]
\centering
\includegraphics[width=0.49\textwidth]{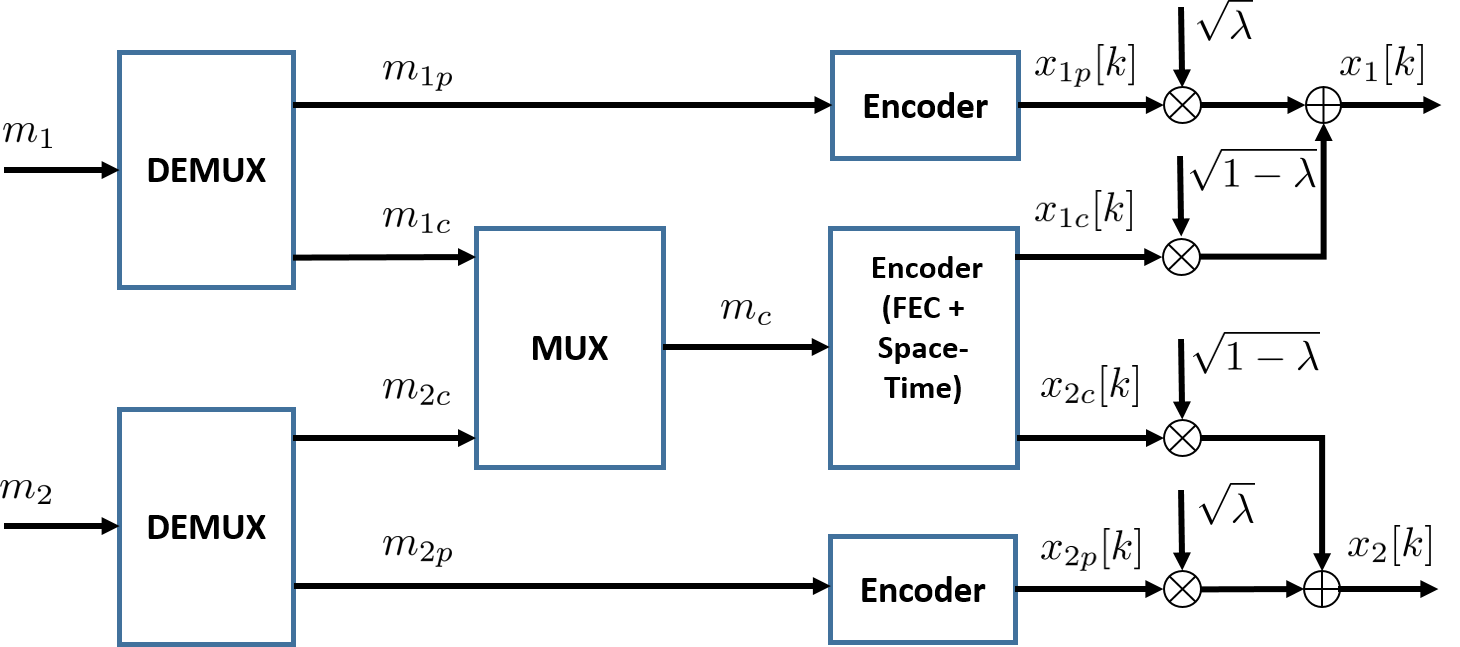}
\caption{Non-Coherent Rate Splitting (NCRS) scheme. DEMUX (MUX)  denotes splitting (combining) of the incoming message(s).}
\label{fig:ncrs}
\end{figure}

The proposed scheme in this paper, which will be termed as Non-Coherent Rate Splitting (NCRS), applies a rate splitting construction, and allows simple analysis and operation with magnitude-only CSIT. 
Both messages are separated into public (or common)  and private sub-messages;  the knowledge of the  channel quality information is exploited to allocate the power and rates to the different sub-messages:  see Fig. \ref{fig:ncrs} for a schematic description of the NCRS message encoding process. 

Due to the complete absence of phase information, NCRS departs from other solutions which operate with  partial CSIT knowledge.
The scheme in  \cite{JafarTIT2017} transmits the same common message from both antennas, in what is known as interference enhancement. It is also different from
\cite{ClerckxTCOM2016}, where the common message is transmitted from all the antennas, after pre-multiplication of the common symbol sequence by different antenna weights.
It is worth mentioning that works such as \cite{ClerckxTCOM2016}, where beamforming and common information decoding are combined,
are usually subject to a global power constraint. 
In Fig.  \ref{fig:ncrs}, messages $(m_1,m_2)$ addressed to the first and second terminals, respectively, are split as 
$m_1 = (m_{1_c},m_{1_p})$ and $m_2 = (m_{2_c},m_{2_p})$. The $m_{1_p}$ and $m_{2_p}$ messages are private messages to be decoded only by user 1 and user 2, respectively, whereas $m_c = (m_{1_c},m_{2_c})$ acts as a public message to be decoded  by both users. 


The messages $m_{1_p}$, $m_c$ and $m_{2_p}$ are encoded and transmitted by the sequences $x_{1p}[k]$, $x_{1c}[k]$, $x_{2c}[k]$ and  $x_{2p}[k]$ as sketched in Fig. \ref{fig:ncrs}, which for a given time instant $k$ read as:
\begin{IEEEeqnarray}{rCl}
	x_1[k] &=& \sqrt{\frac{P}{2}(1-\lambda_1)} \; x_{1c}[k] + \sqrt{\frac{P}{2} \lambda_1} \; x_{1p}[k], \\
	x_2[k] &=& \sqrt{\frac{P}{2}(1-\lambda_2)} \; x_{2c}[k] + \sqrt{\frac{P}{2}\lambda_2} \; x_{2p}[k].
      \end{IEEEeqnarray}
Each receiver first decodes the common message, cancels it from the received signal and then decodes its respective private message. 

  The relative contribution of the two users to the common message can be modulated, in such a way that the information rates of the two users can be expressed as        
\begin{IEEEeqnarray}{ll}  
 R_1 &  = R_{1p} + \alpha \cdot R_c, \\
 R_2 &  = R_{2p} + (1-\alpha) \cdot R_c,
\end{IEEEeqnarray}
with the sum-rate equal to $R_{1p}+R_{2p}+R_c$. $R_{1p}, R_{2p}$ and $R_c$ denote the information rates of the two private messages and the common message, respectively,  with the relative contribution of the two messages to $R_c$ modulated by $0\leq \alpha \leq 1$.

If all submessages are encoded into Gaussian codebooks, then the achievable rate region of NCRS is the convex hull of the regions
$\{\mathcal{R}(\lambda_1,\lambda_2,\alpha),0\leq \lambda_1, \lambda_2, \alpha \leq 1\}$, given by
{\small
\begin{align}\label{eq:R}
 	\mathcal{R}(\lambda_1,\lambda_2,\alpha)   & = 	\Biggl\{  &   \nonumber \\ \nonumber
	R_{1}  &  < \log_2\left(1+\frac{\lambda_1\gamma_{11}}{1+\lambda_2\gamma_{12}}\right) + \alpha \cdot R_c  & \\ \nonumber
		R_{2}  &<   \log_2\left(1+\frac{\lambda_2\gamma_{22}}{1+\lambda_1\gamma_{21}}\right)  + (1-\alpha) \cdot R_c  & \\ \nonumber
		 R_c & = \min \Biggr( \log_2\left(1+\frac{(1-\lambda_1) \gamma_{11}+(1-\lambda_2) \gamma_{12}}{1+\lambda_1\gamma_{11}+\lambda_2\gamma_{12}}\right), & 
	\\ & \log_2\left(1+\frac{(1-\lambda_1) \gamma_{21}+(1-\lambda_2) \gamma_{22}}{1+\lambda_1\gamma_{21}+\lambda_2\gamma_{22}}\right)\Biggr) \Biggr\}.  &
\end{align}}%
The first term of both $R_1$ and $R_2$ in (\ref{eq:R}) correspond to the respective private rates, limited by the interference caused by the private symbols addressed to the other user. The common message is removed after being decoded by both terminals. Its achievable rate is set by the most restrictive terminal under the presence of the private messages, this is why the minimum of the rate of two $2\times 1$ links needs to be taken in (\ref{eq:R}).

Given the lack of phase information, the transmitter is unable to discriminate among different spatial directions. 
The multicast rate of the common message in \eqref{eq:R} can be achieved with Alamouti encoding, which transforms the vector channel into a scalar channel. Thus,
scalar codes can be used to attain the no CSIT MISO channel capacity for the two-antenna case \cite{jafar2005}.
The available magnitude information is used to determine how much private and common information is sent from each antenna when optimizing  the rates in \eqref{eq:R}.
As exposed earlier, NCRS uses less information than DPC and other linear precoding schemes at the transmit side,
and it can be seen that has lower complexity than  HK's joint decoding at the receive terminals.  

\vspace*{-0.3cm}

\subsection{Symmetric Case}\label{sec:optimization}

In order  to get some useful insight, we address the symmetric case, with $\snr = \snr_1=\snr_2$ and $\inr = \inr_1=\inr_2$ in \eqref{eq:snrs}.  
We need to determine only  one variable $\lambda = \lambda_1 = \lambda_2$ to characterize identical rates for both users.  From (\ref{eq:R}), the NCRS sum-rate reads as 
\begin{multline}
  \label{eq:Rsym}
  R = R_c + 2R_p = \log_2\left(1+\frac{(1-\lambda)(\snr+\inr)}{1+\lambda(\snr+\inr)}\right) + \\
  2\log_2\left(1+\frac{\lambda \snr}{1+\lambda \inr}\right),
  \end{multline}
which is maximum for 
\begin{equation}
  \label{eq:lambdasym}
  \lambda^\star = \min\left\{\frac{\snr - \inr}{\inr(\snr+\inr)},1\right\},
\end{equation}
provided that $\snr > \inr$. 
This rate is necessarily not worse than that for the HK scheme, when both transmitters cannot  cooperate. We will show the comparison for a specific case in the next section; 
the expression of the  corresponding HK lower bound for the sum-rate in the symmetric case can be seen in  \cite[Ex. 6.16]{ElGamal2011}, and can be derived by noting that the rate splitting is such that the public messages sent from both antennas  give rise to two Multiple-Access Channels (MAC). 


%

  \section{Numerical results}\label{sec:results}
\begin{figure}
    \centering
  \subfloat[\label{fig:symmetrica}]{%
       \includegraphics[width=0.7\linewidth]{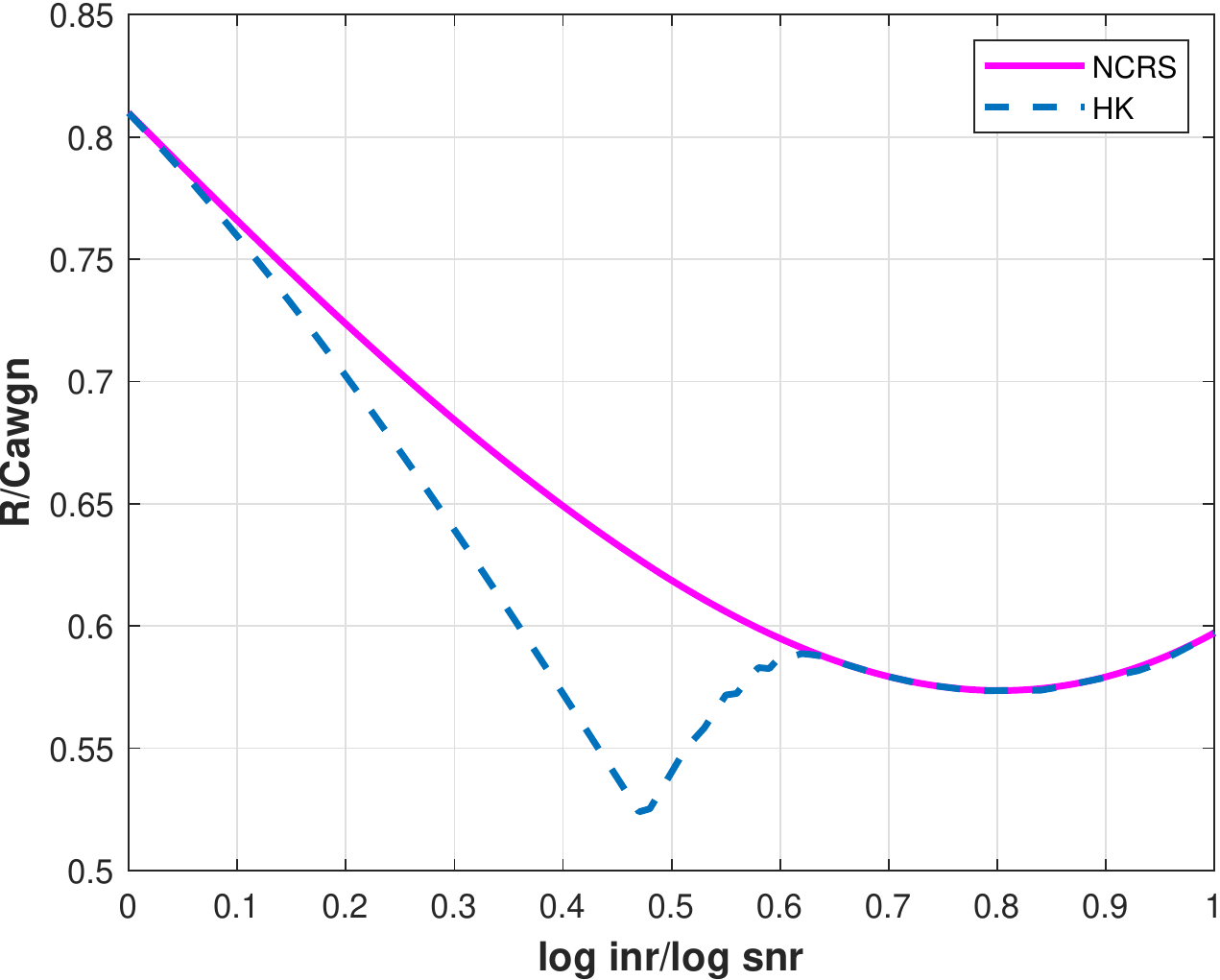}}
    \\
  \subfloat[\label{fig:symmetricb}]{%
        \includegraphics[width=0.7\linewidth]{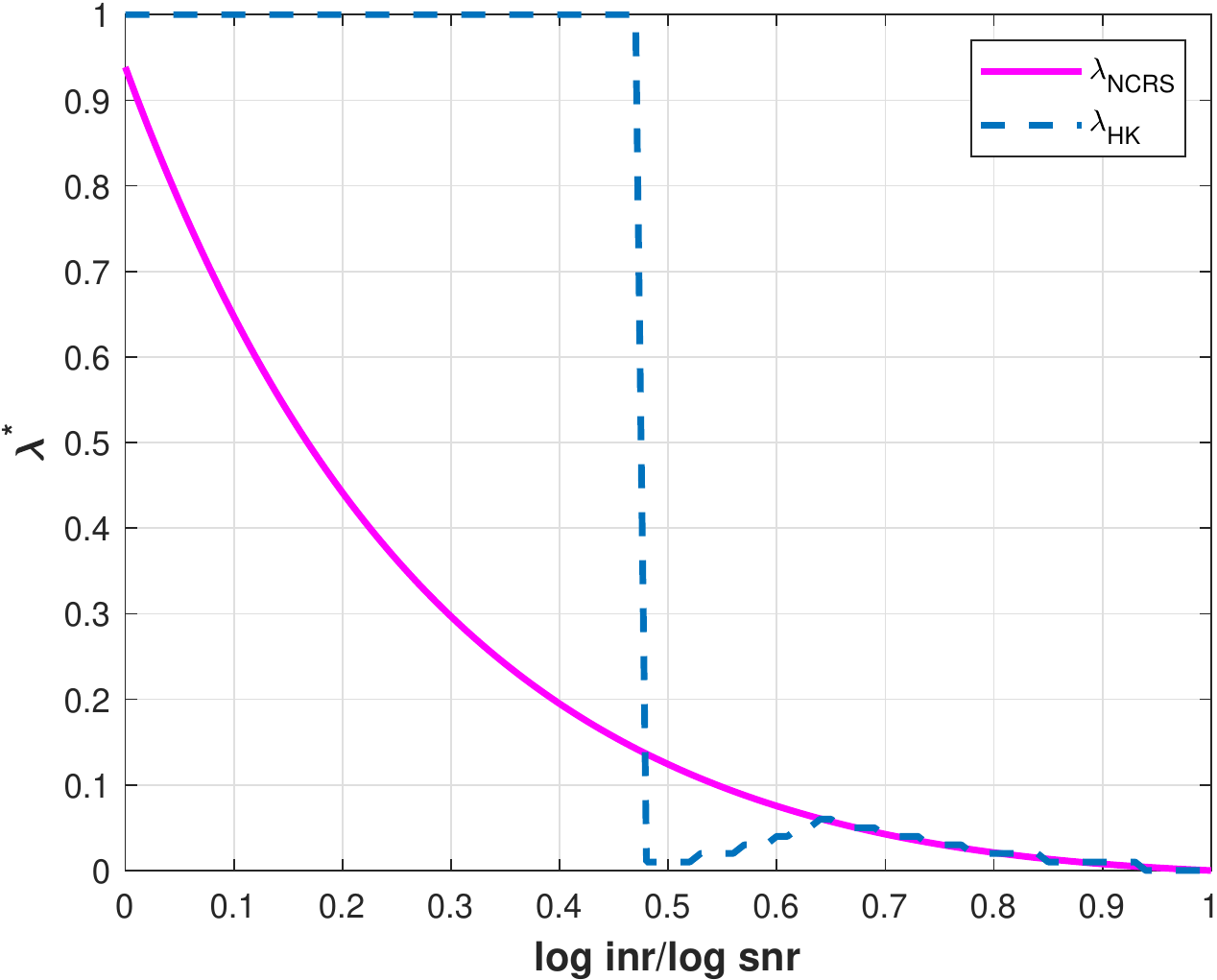}}
  \caption{Comparison of NCRS and HK in the symmetric case, $\snr = 15$ dB. (a) Sum-rate. (b) Weighting factor.}
  \label{fig:symmetric} 
\end{figure}
\begin{figure*}[!tpbh]
\centering
\includegraphics[width=0.95\textwidth]{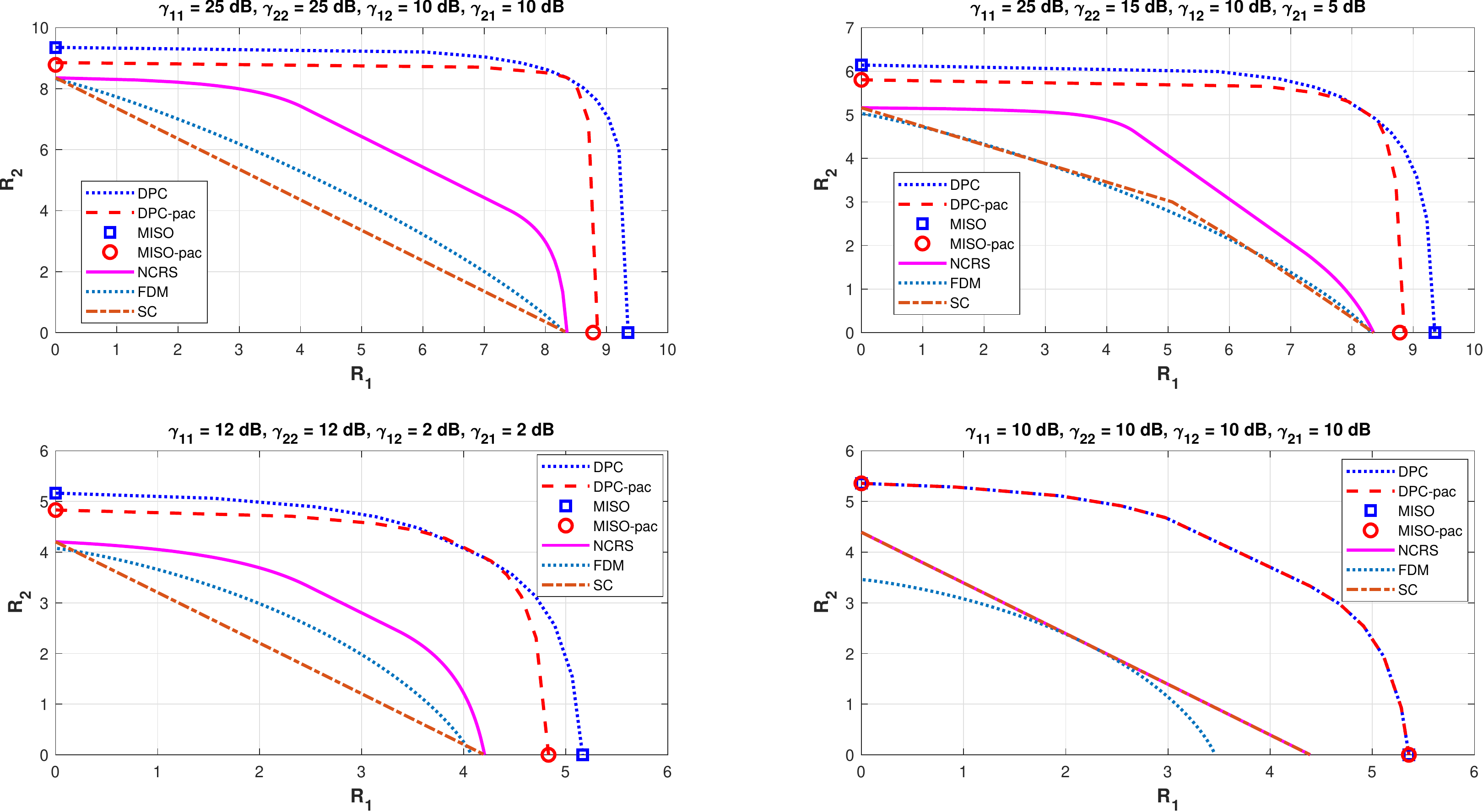}
\caption{Achievable rate regions of  NCRS in contrast to the channel capacity of different baseline schemes.  $R_1$ and $R_2$ are given in bps/Hz. In the bottom right case, NCRS and SC performance are identical, and the same happens with DPC and DPC-pac.}
\label{fig:RateRegions}
\end{figure*}

  First we compare the achievable sum-rate by both NCRS and HK in a symmetric scenario, with results shown in Fig. \ref{fig:symmetric}.
  The ratio of the sum-rates with respect to the sum-rate without interference, $C_{\mathrm{awgn}} = 2\cdot \log_2(1+\snr)$,   is depicted together with the optimal weighting factor $\lambda^*$, with respect to the ratio $\log \inr / \log \snr$.
  The gain of NCRS with respect to HK, which is relevant for intermediate interference regimes,  comes from the cooperation when encoding the two messages $m_1$ and $m_2$ and synthesizing the transmit symbols $x_1[k]$ and $x_2[k]$ in Fig.  \ref{fig:ncrs}.
Next we show the achievable rate region of NCRS  using for comparison the following baseline metrics: (i) DPC \cite{heath2018}; (ii) DPC with per-antenna power constraints (DPC-pac) \cite{yu2007}; (iii) 
Single-user MISO capacity under global power constraint and full CSIT (MISO) \cite{heath2018}; (iv) Single-user MISO capacity under per-antenna power constraint and full CSIT (MISO-pac) \cite{vu2011}; (v) superposition coding (SC) \cite{jafar2005}; and (vi) frequency-division multiplexing (FDM). 
Fig.  \ref{fig:RateRegions} illustrates the results for different channel magnitudes. The gain of NCRS with respect to SC is due to the knowledge of the magnitude of the individual links, which allows to optimize the amount of information coming out of each antenna to a given terminal. Depending on the specific setting, the gap till the channel capacity with full CSIT can be significant, 
and it remains an open problem to determine  whether the rate region achieved by NCRS can be further improved with some alternative encoding scheme.

  


\section{Conclusions}

The two-user vector broadcast channel has been addressed under the absence of phase information at the two-antenna transmitter. A rate splitting approach between private and common messages, properly optimized,
has been used to exploit the knowledge of the magnitude of the individual paths, and compared with previously known  results in the literature which share only some of the constraints of the problem under study. Further work can address a number of users  higher than two, although the complexity of the rate splitting scheme is expected to grow significantly. 


\bibliographystyle{IEEEtran}
\bibliography{DoF}
\end{document}